\documentstyle[preprint,aps]{revtex}
\draft
\begin{document}
\title{Spinor Bose Condensates in Optical Traps}
\author{Tin-Lun Ho}
\address{Department of Physics,  The Ohio State University, Columbus, Ohio
43210}
\maketitle

\begin{abstract}
In an optical trap, the ground state of
spin-1 Bosons such as $^{23}$Na, $^{39}$K, and $^{87}$Rb can be either
a ferromagnetic or a ``polar" state, depending on the scattering lengths in
different angular momentum channel. The collective modes of these states
have very different spin character and spatial distributions.
While ordinary vortices are stable in
the polar state, only those with unit circulation are stable in
the  ferromagnetic state.
The ferromagnetic  state also has coreless (or Skyrmion) vortices like
those of superfluid $^{3}$He-A.
\end{abstract}

Two months ago, the MIT group has succeeded in trapping a
$^{23}$Na Bose condensate by purely optical means\cite{optical}. This
experiment has opened up a new direction in the study of confined dilute atomic
gases. In conventional magnetic traps, the spins of the alkali atoms are
frozen. As a result, even though the alkali atoms carry spins, they behave like
scalar particles. In contrast, the spin of the alkali
atoms are essentially free in an
optical trap. The spinor nature of alkali Bose condensate can
therefore  be manifested\cite{spingauge}.
Because of the wide range of hyperfine spins of the
alkali Bosons and Fermions, the optical trap has provided great
opportunities to study dilute quantum gases of atoms with large spins.

The purpose of this paper is to point out the general properties of
the spinor Bose condensates. As we shall see, they possess
a whole host of quantum phenomena
absent in scalar condensates. These include changes in
ground state structures with interaction parameters,
vector and quadrupolar spin wave modes, topological and energetic instability
of doubly quantized singular vortices, and the existence of coreless
(or Skyrmion) vortices. All these results
are simple consequences of the effective low energy Hamiltonian of the system.
The derivation of this Hamiltonian (which applies to Bosons and
Fermions with arbitrary spins) and the realization of its limitations
are therefore crucial for our discussions.

For simplicity, we shall consider Bosons with hyperfine spin $f=1$.
This includes alkalis with nuclear
spin $I=3/2$ such as $^{23}$Na, $^{39}$K, and $^{87}$Rb.  Alkali Bosons with
$f>1$ such as $^{85}$Rb (with $I=5/2$) and $^{133}$Cs (with
$I=7/2$) have even richer structures and will be discussed elsewhere.
To illustrate the fully degenerate spinor nature, and as a first step,
we shall only consider the
case of zero magnetic field. To reach a good approximation of this limit,
the Zeeman energy must be much smaller than the interaction energy at the
center of the trap (which is essentially the chemical potential $\mu$).
The condition is therefore $\gamma B<<\mu$, where $\gamma$ is the gyromagnetic
ratio, and $B$ is the magnetic field. For clouds with $10^{6}$ atoms,
with scattering length $a_{sc}=100\AA$ and trap frequency $200$Hz,
this condition gives $B<<10^{-3}$gauss. This bound can be further increased by
increasing the trap frequency. Since the current
capability of magnetic shielding
can reach $10^{-5}$gauss, a good approximate of the zero field limit is
attainable.

Our key results are  :
\noindent {\bf (1)} The spinor ground state
depends crucially on the s-wave scattering lengths $\{ a_{F}\}$
of various elastic scattering channel with total hyperfine spin $F$.
When $a_{0}>a_{2}$, the ground state is a
ferromagnet denoted by a normalized spinor $\zeta^{+}= (1,0,0)$.
When $a_{2}>a_{0}$, the ground state is a
 non-magnetic state ``polar " state denoted by
a normalized spinor $\zeta^{+}= (0,1,0)$.
Current estimates of $a_{0}$ and
$a_{2}$ indicate that the spinor condensates of
$^{23}$Na and
$^{87}$Rb may be a polar and a ferromagnetic state respectively.

\noindent {\bf (2)} The
polar state has a density mode ($\omega^{p}_{0}$)
and two degenerate ``vector-like" spin wave modes ($\omega^{\em p}_{\pm}$), all
of which are of Stringari form\cite{Stringari}.
For large clouds, the density
mode is universal (i.e. independent of interaction), whereas
$\omega_{\pm}=\frac{a_{2}-a_{0}}{2a_{2}+a_{0}}\omega_{0}$.
The ferromagnetic state has a density mode similar to that of the polar state.
It also has a ``vector-like" and a ``quadrupolar-like" spin wave mode,
corresponding spin flips of one and two units. Both spin waves modes
obey a Schrodinger equation and reside near the surface of the cloud.
The quadrupolar modes is separated from the vector mode by a gap of the order
of the trap frequency.

\noindent {\bf (3)} In case of ferromagnetic states, vortices with even
circulation ($\pm 2m\pi$, $m>1$) are neither topological or energetically
stable.
Only vortices with unit circulation ($\pm 2\pi$) have topological
stability, a phenomena identical to that of superfluid $^{3}$He-A.

{\em Optical Trap Potential and Their Spin Dependence}: We begin by examining
 the
confining potential produced by a laser with a linearly polarized electric
field ${\cal E}(t)\hat{\bf x}$ and frequency $\omega$.
The potential seen by an atom in hyperfine state
$|1,m>$ is $U_{m} =  -\frac{1}{2}\alpha_{m}\overline{{\bf \cal{E}}^2}$,
\begin{equation}
\alpha_{m} = \sum_{\ell} e^2\frac{|X_{\ell m}|^2 \omega_{\ell m} }
{\omega_{\ell m}^{2} -\omega^{2}}, \,\,\,\,\,\,\,
\sum_{\ell} |X_{\ell m}|^2
\omega_{\ell m} = \frac{\hbar}{2M}Z.
\label{Um} \end{equation}
where $\overline{(..)}$ denotes the time average, $\alpha_{m}$ is the
polarizability of the state $|1,m>$, $\hbar\omega_{\ell m}>0$
is the excitation energy from the ground state $|1,m>$ to excited
states $|\ell>$,  $eX_{\ell m}$ is the dipole matrix element between
$|1,m>$ and $|\ell>$,  and $Z$ is the
atomic number.
We have also written down the dipole sum rule in eq.(\ref{Um}).
For a red detuned laser, $\omega <\omega_{\ell m}$, the atom will be
attracted to region of strong field. A focused laser beam will therefore form
a harmonic trapping potential near the point of maximum intensity.

In general, the spin dependence of the trapping potential is weak.
The reason is that to one part in $10^2$ to $10^3$, the dipole sum rule
is saturated by the $nS$ to $nP$ transitions\cite{saturation},
where $nS$ is the ground state
electronic configuration. The sum rule in eq.(\ref{Um})
can therefore be approximated as
$\sum_{\ell\epsilon P} |X_{\ell m}|^2 \omega_{\ell m} = \frac{\hbar}{2M}Z$.
In addition, the excitation energies
$\omega_{\ell m}$ for different $nS$ to $nP$ transitions differ from each
other only by the fine splitting, which is typically $10^{-3}$ smaller than the
energy difference ($\hbar\omega_{PS}$) between $nS$ to $nP$ levels before
they fine split. Thus, for a far detuned laser, eq.(\ref{Um}) can be
approximated as
\begin{equation}
\alpha_{m} = -\frac{1}{\omega_{PS}^{2}-\omega^{2}}
\left(\sum_{\ell\epsilon P}|ex_{\ell m}|^{2} \omega_{\ell m}\right)
+O\left(\frac{\Delta_{\rm fine}}{\Omega}\right),
\label{Usym} \end{equation}
where $\Delta_{\rm fine}$ is the fine splitting and $\Omega\equiv
\omega_{PS}-\omega$ is the detuning frequency.
To the extent that the transitions $nS\rightarrow n'P$ ($n'>n$)
and the ratio $\Delta_{fine}/\Omega$ can be ignored, the sum
in the first term in Eq.(\ref{Usym}) becomes a constant and the correction
in the second term can be set to zero. This implies that $\alpha_{m}$,
 hence the trapping potential $U_{m}$,
is independent of $m$.

{\em Effective low energy Hamiltonian} : Alkali atoms (Bosons or Fermions)
have two hyperfine multiplets (denoted as $f_{high}$ and $f_{low}$), with
energy difference many orders of magnitude
larger than the trap frequency.
Since the interaction between two alkali atoms depends on their electron spins
(singlet or triplet), the hyperfine states of the atoms can be changed
after the scattering.
However, when the system is very cold, two atoms in  $f_{low}$
will remain in the same multiplet after the scattering since there is
not enough energy to promote either atom to $f_{high}$.
In contrast, energy conservation does not limit the production of $f_{low}$
states
from the scattering of two $f_{high}$ states.  As a result, in an optical
trap, all atoms in the ground state will be in the lower multiplet.
The low energy dynamics of the system will then be described by
a pairwise interaction (between atoms 1 and 2) that preserves
the hyperfine spin of the individual atoms and is rotationally invariant
in the hyperfine spin space. The general form of this interaction is
\begin{equation}
\hat{V}({\bf r}_{1}-{\bf r}_{2}) = \delta( {\bf r}_{1}-{\bf r}_{2})
\sum_{F=0}^{2f} g_{F}{\cal P}_{F},
\,\,\,\,\, g_{F}= 4\pi\hbar^2 a_{F}/M
\label{V} \end{equation}
where $M$ is the mass of the atom,
${\cal P}_{F}$ is the projection operator which projects the pair 1 and 2
into a total hyperfine spin $F$ state, and $a_{F}$
the s-wave scattering length in the total spin $F$ channel.
For Bosons (or Fermions), symmetry implies that only even (or odd) $F$ terms
appear in eq.(\ref{V}).

For a system of $f=1$ Bosons, we have
$V = g_{2}{\cal P}_{2} + g_{0}{\cal P}_{0}$. Likewise, the relation
${\bf F}_{1}\cdot{\bf F}_{2}$$=$$\sum_{F=0}^{2f}$
$\lambda_{F}$${\cal P}_{F}$, $\lambda_{F}\equiv
\frac{1}{2}\left[F(F+1)-2f(f+1)\right]$ becomes
${\bf F}_{1}\cdot{\bf F}_{2} = {\cal P}_{2} - 2{\cal P}_{0}$.
The relation $\sum_{F}{\cal P}_{F=0}^{2f}=1$ becomes
$1 = {\cal P}_{2}+{\cal P}_{0}$. We then have  (dropping the
$\delta$-function in eq.(\ref{V})),
\begin{equation}
V = c_{0} + c_{2} {\bf F}_{1}\cdot{\bf F}_{2},
\end{equation}
where $c_{0} = (g_{0}+2g_{2})/3$ and $c_{2} = (g_{2}-g_{0})/3$.
The Hamiltonian in the second quantized form
is then
\begin{equation}
{\cal H} = \int {\rm d}{\bf r}\left(
\frac{\hbar^2}{2M}{\bf \nabla} \psi^{+}_{a}\cdot {\bf \nabla}\psi_{a}  +
U\psi^{+}_{a}\psi_{a}  +
\frac{c_{0}}{2}\psi^{+}_{a}
\psi^{+}_{a'}\psi_{a'}\psi_{a}
+ \frac{c_{2}}{2}\psi^{+}_{a}\psi^{+}_{a'}
{\bf F}_{ab}\cdot{\bf F}_{a'b'}\psi_{b'}\psi_{b}\right)  \label{H}
\end{equation}
where $\psi_{a}({\bf r})$ is the field annihilation operator for an atom
in hyperfine state $|1,a>$ at point ${\bf r}$, ($a=1,0,-1$), and $U$ is the
trapping potential.
It is straightforward to generalize eq.(\ref{H}) to higher spins.
By noting that $({\bf S}_{1}\cdot {\bf
S}_{2})^{n}=\sum_{F=0}^{2f}\lambda^{n}_{F}{\cal P}_{F}$,
eq.(\ref{V}) can be
written as $V= \sum_{n=0}^{f} c_{n}({\bf F}_{1}\cdot{\bf F}_{2})^{n}$,
where $c_{n}$ is a linear combination of the $g_{F}$'s.
Alternatively, it is useful to express the projection operator
${\cal P}_{F}$ in the second quantized form as
${\cal P}_{F}= \sum_{a=-F}^{F}\hat{O}^{+}_{Fa}\hat{O}_{Fa}$,
where $\hat{O}_{Fa}$$=$
$\sum_{a_{1},a_{2}}<Fa|f,a_{1};f,a_{2}> \hat{\psi}_{a_{1}}\hat{\psi}_{a_{2}}$,
and $<Fm|f,a_{1}; f,a_{2}>$ is the Clebsch-Gordon coefficient for forming a
total spin $F$ state from two spin-$f$ particles.
The pair potential is then
$\hat{V}=\sum_{F=0}^{2f} g_{F}\sum_{a=-F}^{F} \hat{O}^{+}_{Fa}\hat{O}_{Fa}$.
Only even and odd $F$ terms appear in $\hat{V}$ in case of
Bosons and Fermions.

{\em Ground State Structure} : It is convenient to write the Bose condensate
$\Psi_{a}({\bf r}) \equiv<\hat{\psi}_{a}({\bf r})>$ as
$\Psi_{a}({\bf r})= \sqrt{n({\bf r})}
\zeta_{a}({\bf r})$, where $n({\bf r})$ is the
density, and $\zeta_{a}$ is a normalized spinor $\zeta^{+}\cdot\zeta =1$.
The ground state structure of $\Psi_{a}({\bf r})$ is
determined by minimizing the energy with fixed particle number, i.e.
$\delta K = 0$, $K\equiv \delta <H-\mu N>$, where $\mu$ is the
chemical potential,
\begin{equation}
K = \int {\rm d}{\bf r} \left( \frac{\hbar^{2}}{2M}({\bf \nabla}\sqrt{n})^2
+ \frac{\hbar^{2}}{2M}({\bf \nabla}\zeta)^2 n -\left[\mu - U({\bf r})
\right] n
+ \frac{n^2}{2}\left[c_{0} + c_{2} <{\bf F}>^2 \right]
\right), \label{K} \end{equation}
and $<{\bf F}>\equiv \zeta^{\ast}_{a}{\bf F}_{ab}\zeta_{b}$.
It is obvious that
all spinors related to
each other by gauge transformation $e^{i\theta}$ and spin rotations
${\cal U}(\alpha, \beta, \tau)$$=$$e^{-iF_{z}\alpha} e^{-iF_{y}\beta}
e^{-iF_{z}\tau}$ are degenerate, where $(\alpha, \beta, \tau)$ are the
Euler angles. There are two distinct cases :

\noindent {\bf I} : Polar state : This state emerges when $c_{2}>0$ (i.e.
$g_{2}>g_{0}$). The energy is minimized by $<{\bf F}>=0$. The
spinor $\zeta$ and the density $n^{o}$ in the ground state are given by
\begin{equation}
\zeta = e^{i\theta} {\cal U} \left( \begin{array}{c} 0\\1\\0\end{array} \right)
= e^{i\theta} \left( \begin{array}{c}
-\frac{1}{\sqrt{2}}e^{-i\alpha}{\rm sin}\beta \\
{\rm cos}\beta  \\ \frac{1}{\sqrt{2}}e^{i\alpha}
{\rm sin}\beta \end{array} \right)
\,\,\,\,\,\,\,\
n^{o}({\bf r}) = \frac{1}{c_{o}}\left(\mu - U({\bf r}) -  W({\bf r})\right)
\label{polardensity} \end{equation}
where $W({\bf r}) = \frac{\hbar^2}{2M}
\frac{{\bf \nabla^2}\sqrt{n^{o}}}{\sqrt{n^{o}}}$. Note that $\zeta$ is
independent of the Euler angle $\tau$. The symmetry group of polar state is
therefore $U(1)\times S^{2}$, where $U(1)$ denotes the phase angle $\theta$,
and $S^{2}$ is a surface of a unit sphere denoting all orientations $(\alpha,
 \beta)$
of the spin quantization axis.

\noindent {\bf II} : Ferromagnetic state : This state emerges when
$c_{2}<0$, or $g_{0}>g_{2}$. The energy is minimized by making
$<{\bf F}>^2=1$. The ground state spinor and density are
\begin{equation}
\zeta = e^{i\theta} {\cal U}
\left( \begin{array}{c} 1 \\0\\0\end{array} \right)
= e^{i(\theta-\tau)} \left( \begin{array}{c}
e^{-i\alpha}{\rm cos}^{2}\frac{\beta}{2} \\
\sqrt{2} {\rm cos}\frac{\beta}{2}{\rm sin}\frac{\beta}{2}
 \\ e^{i\alpha}{\rm sin}^{2}\frac{\beta}{2} \end{array} \right)
\,\,\,\,\,\,\,\,
n^{o}({\bf r}) = \frac{1}{g_{2}}\left(\mu - U({\bf r}) -  W({\bf r})\right)
\label{ferrodensity} \end{equation}
The direction of the spin is $<{\bf F}>= {\rm cos}\alpha \hat{\bf z}$$+$
${\rm sin}\alpha ({\rm cos}\beta \hat{\bf x} + {\rm sin} \beta \hat{\bf y})$.
The combination $(\theta-\gamma)$ in eq.(\ref{ferrodensity}) clearly displays
a ``spin-gauge" symmetry \cite{spingauge},
i.e. the equivalence between phase change and spin rotation.
Because of this symmetry,
the distinct configurations of $\zeta$ (including the gauge)
are given by the full range of the
Euler angles. The symmetry group is therefore $SO(3)$.
As we shall see, this difference in symmetry between the polar and the
ferromagnetic state leads to a fundamental difference in their vortices.

According to the latest estimate of J. Burke, J. Bohn, and
C. Greene\cite{private1}, the scattering of $^{23}$Na are
$a_{2}= (52\pm 5)a_{B}$ and $a_{0}= (46\pm 5)a_{B}$; and those for
for $^{87}$Rb are $a_{2}= (107\pm 4)a_{B}$ and $a_{0}= (110\pm 4)a_{B}$,
where $a_{B}$ is the Bohr radius. Because of the overlapping error bars of
$a_{2}$ and $a_{0}$ in each case, one can not say for sure about
the nature of their ground states. However, if the inequalities suggested by
current estimate ($a_{2}>a_{0}$ for $^{23}$Na and $a_{0}>a_{2}$
for $^{87}$Rb)  are true, then the condensates of $^{23}$Na and
 $^{87}$Rb are polar state and  ferromagnetic state respectively.

{\em Collective modes of trapped spinor Bose condensates}:
The equation of motion in zero field is
\begin{equation}
i\hbar \partial_{t}\hat{\psi}_{m}= -\frac{\hbar^{2}}{2M}{\bf \nabla}^2
\hat{\psi}_{m} + [U({\bf r})-\mu]\hat{\psi}_{m}
+ c_{0}\left(\hat{\psi}^{+}_{a}\hat{\psi}_{a}\right)\hat{\psi}_{m}
+ c_{2}\left(\hat{\psi}^{+}_{a}{\bf F}_{ab}\hat{\psi}_{b}\right)
\cdot \left({\bf F}\hat{\psi}\right)_{m}
\label{motion} \end{equation}
To study the elementary excitations, we write
$\hat{\psi}_{m}=\Psi^{o}_{m}+\hat{\phi}_{m}$ and linearizing eq.(\ref{motion})
about the ground state $\Psi^{o}$.

\noindent {\bf I} : Polar State : Without loss of generality, we take
$\hat{\bf z}$ as  the spin quantization axis and
$\zeta^{T}=(0,1,0)$, with the subscript ``$T$" denotes the transpose.
Using the expression of $n^{o}$ in eq.(\ref{polardensity})
and the fact that $<{\bf F}>=0$,
eq.(\ref{motion}) becomes
\begin{equation}
i\hbar \partial_{t}\left(\begin{array}{c}\hat{\phi}_{0}\\ -\hat{\phi}_{0}^{+}
\end{array}\right) = - \frac{\hbar^{2}}{2M}{\bf \nabla}^2
\left(\begin{array}{c} \hat{\phi}_{0}\\ \hat{\phi}_{0}^{+} \end{array}\right)
+ W({\bf r})
\left(\begin{array}{c} \hat{\phi}_{0}\\ \hat{\phi}_{0}^{+} \end{array}\right)
+ n^{o} c_{0}
\left(\begin{array}{c}\hat{\phi}_{0} + \hat{\phi}^{+}_{0} \\
\hat{\phi}_{0} + \hat{\phi}^{+}_{0}
\end{array}\right)
\label{pdensity} \end{equation}
\begin{equation}
i\hbar \partial_{t}\left(\begin{array}{c}\hat{\phi}_{1} \\
-\hat{\phi}_{-1}^{+} \end{array}\right)
= - \frac{\hbar^{2}}{2M}{\bf \nabla}^2
\left(\begin{array}{c} \hat{\phi}_{1}\\ \hat{\phi}_{-1}^{+} \end{array}\right)
+ W({\bf r})
\left(\begin{array}{c} \hat{\phi}_{1}\\ \hat{\phi}_{-1}^{+} \end{array}\right)
+ n^{o} c_{2}
\left(\begin{array}{c}\hat{\phi}_{1} + \hat{\phi}^{+}_{-1} \\
\hat{\phi}_{1} + \hat{\phi}^{+}_{-1}
\end{array}\right)
\label{pspin} \end{equation}
To linear order in $\hat{\phi}_{m}$, the density and spin
fluctuations ($\delta \hat{n}$ and $\delta \hat{M}_{\pm}$) are related to
$\hat{\phi}_{0}$ and $\hat{\phi}_{\pm}$ as
$\delta \hat{n}({\bf r}) = \sqrt{n^{o}({\bf r})} \left(
\hat{\phi}_{0}+\hat{\phi}_{0}^{+}\right) $,
$\delta \hat{M}_{+} \equiv \delta (\hat{M}_{x} + i\hat{M}_{y})$
$= \sqrt{n^{o}({\bf r})} \left(
\hat{\phi}_{1}+\hat{\phi}_{-1}^{+}\right) $, and $\hat{M}_{-}=\hat{M}_{+}^{+}$.
Denoting the frequencies of $\hat{\phi}_{0}$ and $\hat{\phi}_{\pm}$
as $\omega_{0}$ and $\omega_{\pm}$, it is easy to see that they are all of
the Bogobuibov form in the homogenous case ($W=0$), where
$\hbar \omega_{0} = \sqrt{\epsilon_{\bf k}(\epsilon_{\bf k} +
2c_{0}n^{o})}$,
$\omega_{\pm } = \sqrt{\epsilon_{\bf k}(\epsilon_{\bf k} +
2c_{2}n^{o})} $, and $\epsilon_{\bf k}= \hbar^2k^2/(2M)$.

In a harmonic trap, eq.(\ref{pdensity}) is identical to the equation of the
collective mode of a scalar Boson. Using the method of
Stringari\cite{Stringari}, it is straightforward to show that for
large cloud (where $n^{o}$ in eq.(\ref{polardensity})
is well approximated by the Thomas Fermi expression
$n^{o}({\bf r})=[\mu - U({\bf r})]/c_{0}$ by ignoring  $W({\bf r})$
\cite{Baym}),  eqs.(\ref{pdensity})
and (\ref{pspin}) can be written as
\begin{equation}
\partial_{t}^{2}\delta \hat{n} = {\bf \nabla} \left(c_{0}n^{o}
{\nabla} \delta n\right), \,\,\,\,\,\,\,
\partial_{t}^{2}\delta \hat{M}_{\pm} = {\bf \nabla} \left(c_{2}n^{o}
{\nabla} \delta \hat{M}_{\pm}\right). \,\,\,\,\,\,\,
\label{Stringari} \end{equation}
Stringari has shown that the density mode in eq.(\ref{Stringari})
has a universal spectrum (i.e. interaction independent) with power law
(hence extended) wavefunctions\cite{Stringari}.
Since the spin waves modes $\delta M_{\pm}$ obey exactly the same equation as
$\delta n$ except that $c_{0}$ is replaced by $c_{2}$, the quantum numbers
and the wave functions of the spin wave modes are identical to those
of the density mode, and
\begin{equation}
\omega_{\pm} = \frac{c_{2}}{c_{0}}\omega_{0}
= \frac{a_{2}-a_{0}}{2a_{2}+a_{0}}\omega_{0}.
\label{spinfre} \end{equation}

\noindent {\bf II} : Ferromagnetic State : We shall take $\zeta^{T}=(1,0,0)$.
Using eq.(\ref{ferrodensity})
and the fact that $<{\bf F}>=\hat{\bf z}$,
eq.(\ref{motion}) becomes
\begin{equation}
i\hbar \partial_{t}\left(\begin{array}{c}\hat{\phi}_{1}\\
\hat{\phi}_{0}\\ \hat{\phi}_{-1}
\end{array}\right) = - \frac{\hbar^{2}}{2M}{\bf \nabla}^2
\left(\begin{array}{c}\hat{\phi}_{1}\\
\hat{\phi}_{0}\\ \hat{\phi}_{-1}
\end{array}\right) + W({\bf r}) \left(\begin{array}{c}\hat{\phi}_{1}\\
\hat{\phi}_{0}\\ \hat{\phi}_{-1}
\end{array}\right) +
n^{o} \left(\begin{array}{c} g_{2}(\hat{\phi}_{1} + \hat{\phi}_{1}^{+})\\
0\\ 2|c_{2}| \hat{\phi}_{-1}
\end{array}\right)
\label{fdensity} \end{equation}
To linear order in  $\hat{\phi}_{a}$, the density,
spin, and ``quadrupolar" spin fluctuations are given by
$\delta \hat{n}=\sqrt{n^{o}}(\hat{\phi}_{1}+\hat{\phi}_{1}^{+})$,
$\delta \hat{M}_{-}=\sqrt{n^{o}}\hat{\phi}_{0}^{+}$, and
$\delta \hat{M}_{-}^{2}=2\sqrt{n^{o}}\hat{\phi}_{-1}^{+}$.
The frequencies of these modes will be denoted as
$\omega_{1}, \omega_{0}$, and $\omega_{-1}$ respectively.
In the homogenous case ($W=0$), the density mode has a Bogoluibov
spectrum $\hbar\omega_{0}=\sqrt{\epsilon_{\bf k}^{2} +  2g_{2}n^{o}
\epsilon_{\bf k}^{2}}$,
$\epsilon_{\bf k}= (\hbar k)^2/(2M)$. The spin wave
$\delta \hat{M}_{-}$ has a free particle spectrum
$\omega_{0}= \epsilon_{\bf k}$. The frequency of $\delta \hat{M}_{-}^{2}$
is free particle like with a gap,
$\omega_{0}= \epsilon_{\bf k} + 2|c_{2}|n^{o}$.

In a harmonic trap, the density mode $\delta \hat{n}$
assumes the Stringari form because $\hat{\phi}_{1}$ obeys the same
equation as scalar Bosons. The spectra of both $\hat{\phi}_{0}$ and
$\hat{\phi}_{-1}$ are given by Schrodinger equations with potentials
$W({\bf r})$ and $W({\bf r}) + 2|c_{2}|n^{o}({\bf r})$ respectively.
Outside the cloud, the potentials for both $\hat{\phi}_{0}$ and
$\hat{\phi}_{-1}$ reduce to the harmonic potential $U({\bf r})-\mu$.
For a large cloud, $n^{o}$ is well approximated by the Thomas Fermi expression
$n^{o}=[\mu-U]/g_{2}$ except near the surface of the cloud. Calculating $W$
with $n^{o}$, one finds that it becomes more attractive near the surface.
The low energy modes of $\delta \hat{M}_{-}$ are therefore confined near
the surface. For the quadrupolar spin waves, $\delta \hat{M}_{-}^{2}$,
the low energy modes are even more confined to the
surface because of the additional potential
$2|c_{2}|n^{o}({\bf r})$, which is
much more repulsive than $W$ inside the cloud.
The spectrum of the surface modes of  $\delta \hat{M}_{-}$ and
 $\delta \hat{M}_{-}^{2}$ therefore mimic their homogeneous counterpart,
with the energy of $\delta \hat{M}_{-}^{2}$ shifted up from that of
$\delta \hat{M}_{-}$ by an amount
$\sim |c_{2}|n^{o}\sim (|c_{2}|/g_{2}) \mu =  (|c_{2}|/g_{2})
(R/a_{T})^{2}\hbar \omega_{T}$, where $R$ is the size of the cloud, and
$a_{T}=\sqrt{\hbar/M\omega_{T}}$ is the trap length.

{\em Intrinsic stability of singular vortices with circulation ($\ell>1$) in
the ferromagnetic state}: The fundamental difference between the
vortices of the polar and the ferromagnetic states can be illustrated by
their superfluid velocities,
${\bf v}_{s}\equiv \frac{\hbar}{M}\zeta^{+}{\bf \nabla}\zeta$. From
eqs.(\ref{polardensity}) and (\ref{ferrodensity}), they are
\begin{equation}
\left({\bf v}_{s}\right)_{polar}=\frac{\hbar}{M}{\bf \nabla} \theta ,
\,\,\,\,\,\,
\left({\bf v}_{s}\right)_{ferro} = \frac{\hbar}{M}
\left( {\bf \nabla}(\theta - \tau) - {\rm cos}\beta{\bf \nabla}\alpha \right),
\label{vs} \end{equation}
where we have assumed that the Euler angles $(\alpha, \beta, \tau)$ and
$\theta$ are spatially varying functions.
Unlike the polar state, the superfluid velocity ${\bf v}_{s}$ of the
ferromagnetic state depends on spin rotations.
This leads to the following remarkable property of the Bose ferromagnets :
if too much vortex energy is stored in one spin component, the system
can get rid of it by spin rotation.
To illustrate this phenomenon, consider the following family of spinor
states $\{ \Psi_{a}(t) =
\sqrt{n^{o}}\zeta_{a}(t)\}$ parametrized by a parameter $t$ between 0 and
$1$,
\begin{equation}
\zeta^{T}(t) = \left( e^{i2m\phi}
{\rm cos}^{2}\left(\frac{\pi t}{2}\right),
e^{im\phi}\sqrt{2} {\rm sin}\left(\frac{\pi t}{2}\right)
{\rm cos}\left(\frac{\pi t}{2}\right),
{\rm sin}^{2}\left(\frac{\pi t}{2}\right) \right),
\label{vfamily} \end{equation}
where $m>0$ is an integer, $\phi$ is the azimuthal angle, and $n^{o}$ is the
equilibrium density for the vortex state $\zeta^{T}(t=0)=(e^{2m\phi i},0,0)$
with $\ell=2m$ circulation. As $t$ evolves from $0$ to 1, this $2m\pi$-vortex
evolves $continuously$ to the vortex free state
 $\zeta^{T}(t=1)=(0,0,1)$ with a spin texture
 $<{\bf F}>$$=$${\rm cos}(\pi t)\hat{\bf z}
+ {\rm sin}(\pi t)[{\rm cos}(m\phi) \hat{\bf x}
+ {\rm sin}(m\phi) \hat{\bf y}]$.

Eq.(\ref{vfamily}) shows that the $2m\phi$ vortex is topologically unstable.
It can only be stabilized by energy barriers preventing its collapse.
Such barriers, however, are non-existent.
From eq.(\ref{K}), it is easily shown that (with $c\equiv {\rm
cos}\left(\frac{\pi t}{2}\right), s^{2}=1-c^{2})$,
\begin{equation}
\frac{ {\rm d}K}{{\rm d}t} = - \frac{\pi}{2}
\frac{\hbar^{2}}{2M} \int {\rm d}{\bf r}
n^{o}({\bf r}) |{\bf \nabla}2m\phi|^2|(c + 2c^3) s <0 , \,\,\,\,\,\,\,
\label{dKdt} \end{equation}
for $0<t<1$.
The fact that $K$ decreases monotonically with increasing $t$
shows the absence of energy barriers.

By multiplying eq.(\ref{vfamily}) by $e^{i\phi}$, one also obtains
a family connecting a vortex with  $2m+1$ circulation
$\zeta^{T}(t=0) = e^{i(2m+1)\phi}(1,0,0)$ to a vortex with unit circulation
$\zeta^{T}(t=1) = e^{i\phi}(1,0,0)$. It can be easily shown as
in eq.(\ref{dKdt}) that ${\rm d}K/{\rm d}t <0$ for $m>0$. This
means that
all $(2m+1)\pi$-vortices will collapse into a $2\pi$-vortex.
The $2\pi$-vortex, however, can not be deformed continuously into a uniform
state\cite{homotopy}.

{\em Coreless (or Skyrmion) vortices}: Eq.(\ref{vs}) shows that spin
variations in the ferromagnetic states in general leads to superflows.
To illustrate further, consider the  condensate
$\zeta({\bf r})^{T}$$=$$({\rm cos}^{2}\frac{\beta}{2},$
$\sqrt{2} e^{i\phi}{\em sin}\frac{\beta}{2}{\rm cos}\frac{\beta}{2},$
$e^{2i\phi}{\em sin}^{2}\frac{\beta}{2})$ where $\beta=\beta(r)$ is an
increasing function of $r$ starting from $\beta=0$ at $r=0$. The
spin texture and superfluid velocity of this condensate are both cylindrically
symmetric,
$<{\bf F}>$$=$$\hat{\bf z}{\rm cos}\beta$
$+$${\rm sin}\beta$$({\rm cos}\phi \hat{\bf x}+ {\rm sin}\phi \hat{\bf y})$,
and ${\bf v}_{s}=\frac{\hbar}{Mr}(1-{\rm cos}\beta)\hat{\phi}$.
 When $\beta(r)$ reaches $\pi/2$,  ${\bf v}_{s}$ becomes the velocity
field of a singular vortex.  However, the singularity of the usual $2\pi$
vortex is absent because ${\bf v}_{s}$ vanishes instead of diverges at $r=0$.
This phenomena
is identical to that of superfluid $^{3}$He-A, which has exactly the same
angular momentum texture, superfluid velocity, and topological instability
\cite{3HeA}\cite{MerminHo}.
In the case of $^{3}$He-A,
it  is known that external rotations can distort the texture so as to
generate a velocity field ${\bf v}_{s}$ to mimic the external rotation as
closely as possible\cite{3HeA}.
Such textural distortions will occur here for the same energetic reason.
physical  reason.

We have thus established Statements {\bf (1)} to {\bf (3)}.
In current experiments, the condensates are first produced in a magnetic trap
and then loaded into an optical trap. Because of the way they are produced,
the condensates will in general carry a net magnetization.
If the spin relaxation time is sufficiently long, the condensate  will behave
like a ferromagnetic state even its ground state is a polar state.
In this case, the vortex phenomena discussed above will apply.

(Notes added: At the time of submission of this paper, a preprint by T. Ohmi
and K. Machida has just appeared (cond-mat/9803160). These authors have studied
the same model in finite field and in the absence of a trap.
Our zero field results of homogeneous systems agree with each other. )

I would like to thank Wolfgang Ketterle, Dan Stamper-Kurn, Eric Cornell,
Carl Wieman, and Greg Lafayatis for valuable discussions.
Special thanks are due to John Bohn, Jim Burke,
and Chris Greene for  kindly providing me with their estimates
of  scattering lengths.
Part of this work was done during a visit to JILA in Dec 97.
I thank Eric Cornell and Carl Wieman for their hospitality.
This work is supported by a Grant from NASA and partly by NSF Grant
DMR-9705295.


\begin{references}
\bibitem{polar} The name ``polar" is chosen because it is the analog of the
Polar state in superfluid $^3$He.
\bibitem{homotopy} Mathematically, the difference between
the vortices of
the polar and the ferromagnetic state can be traced to their different
first homotopy group,  $\pi_{1}(U(1)\times S^{2})=Z$ and
 $\pi_{1}(SO(3))=Z_{2}$, meaning that the polar phase has infinitely many
distinct defects of different winding, whereas the ferromagnetic phase has only
 one defect with non-trivial winding.
\end{references}
\end{document}